\documentclass[leqno, a4paper]{article}
\usepackage{graphicx}
\usepackage{cases}
\usepackage{epsfig}
\begin{document}
\title{Predicting Failures of Point Forecasts\label{weather}}
\author{\small S.~Hallerberg$^{1,2,5}$, J.~Br\"{o}cker$^{2,3,4}$, H.~Kantz$^{2,4}$, L.~A.~Smith$^{3}$\\[0.2cm]
\small 1 Institute of Physics, Chemnitz University of Technology,\\
\small D-09107 Chemnitz, Germany;\\[0.2cm]
\small 2 Max-Planck-Institute for the Physics of Complex Systems,\\
\small N\"othnitzer Str.\ 38, D 01187 Dresden, Germany;\\[0.2cm]
\small 3 Centre for the Analysis of Time Series.\\
\small Department of Statistics. London School of Economics,\\
\small London WC2A 2AE. UK;\\[0.2cm]
\small 4 Center for Dynamics, TU Dresden, 01062 Dresden, Germany;\\[0.2cm]
\small 5 Max-Planck-Institute for Dynamics and Self-Organization,\\
\small Am Fa{\ss}berg 17, 37077 G{\"o}ttingen, Germany
}
\date{}
%
%\correspondence{{Sarah Hallerberg}\\ (sarah.hallerberg@physik.tu-chemnitz.de)}
%
%\firstpage
\maketitle
%%%%%%%%%%%%%%%%%%%%%%%%%%5
\begin{abstract}
The predictability of errors in deterministic temperature forecasts is investigated.
More precisely, the aim is to issue warnings whenever the differences between forecast and verification exceed a given threshold.
%
%This task can be formulated in terms of a classification problem:
%
The warnings are generated by analyzing the output of an ensemble forecast system in terms of a decision making approach.
The quality of the resulting predictions is evaluated by computing receiver operating characteristics, the Brier score, and the Ignorance score.
Special emphasis is also given to the question whether rare events are better predictable.
\end{abstract}
%%%%%%%%%%%%%%%%%%%%%%%%%%%%%
\section{Introduction}
For many practical applications it is not only of interest to have a forecast of a meteorological variable, but also to have information about the possible deviation of this forecast from the observation.
Examples for such applications could be the estimation of the expected electricity demand \cite{TaylorBuizza2003} or the output of a wind farm.
In both cases, a forecast error estimate would be helpful to reveal possible risk exposures.
Various measures of the forecast skill provide averaged estimates of forecast accuracy. 
However, at a given time the instantaneous accuracy might deviate from the averaged forecast skill, since the former depends on the present state of the atmosphere.

We are interested in predicting large instantaneous differences between a deterministic forecast and the corresponding observation (verification).
Predictions of these differences are obtained by post--processing the output of an ensemble forecast system in terms of a classification or decision making approach.
More precisely, we analyze the present state of the ensemble forecast system in order to decide whether a large difference between forecast and verification is impending and therefore a warning should be issued.
This approach is based on the assumption that the ensemble reflects the uncertainty about the future state of the atmosphere.
In the simplest case, the ensemble can be thought of as a collection of equally likely scenarios of the atmosphere's future development.
Operational ensembles show deviations from this clearly idealistic behavior \cite{broecker06-5}.
The idea of the decision making approach to failures of point forecasts is related to predictions made through the identification of precursory structures in time series.
Observations are compared with structures that are believed to be relevant precursors for an event that is expected to occur in the near future.
Precursor based forecasts are typically used in situations that do not allow for a modeling of the system under study but provide a time series record of the past.
Typical examples for predictions through precursory structures are earthquakes \cite{KossobokovM81984}, epileptic seizures \cite{mormann2003} and predictions of turbulent wind gusts \cite{Physa}.
In this contribution we combine precursor based predictions with deterministic forecasts produced by dynamical atmospheric models.
While the dynamic model 
issues deterministic forecasts of a meteorological variable,
the precursor based approach allows to predict possible failures of these point forecasts.
In this context, we treat the verification, the dynamic model and the corresponding ensemble forecasts as a multivariate time series.
Loosely speaking, the precursors we are looking for live in the subset of the multivariate time series spanned by the output of the ensemble forecast system, and the high resolution forecast.
The events we are aiming to predict live in another subset consisting of the high resolution forecast and the time series of observations.
The dependence between event and precursor 
can then be understood as a consequence of the fact that the high resolution forecast and the ensemble forecasts are supposed to describe the same state of the atmosphere.
In the context of precursor based predictions it has been observed that the quality of the predictions can display a strong dependence on the event magnitude \cite{Physa,Goeber}.
In previous work we studied this dependence of the prediction quality on the event magnitude in more detail by predicting events in one--dimensional stochastic processes, as well as wind speed recordings \cite{Sarah2}.
Consequently, we are now not only interested in predicting large deviations of the forecast but also in studying the dependence of these predictions on the threshold that is used to define the deviations.
In Sec.~\ref{data} we specify the properties of the data record used for this study. 
In Sec.~\ref{events} we define the events of interest and analyze their occurrence in the data set.
In the following section we introduce two different strategies to identify suitable precursors, and develop a corresponding setting for decision making.
In Sec.~\ref{quality} we use these strategies to issue warnings and analyze the quality of these predictions by computing receiver operating characteristics, Brier scores, and Ignorance scores.
Sec.~\ref{magnitude} is devoted to understanding the relation between the quality of the predictions and the magnitude of the events under study.
We summarize in Sec.~\ref{conclusions}.   
%
%
%%%%%%%%%%%%%%%%%%%%%%%%%%%%%%%%%%%%%%%%%%%%%%%%%%%%%%%%%%%%%%%%%%%%
\section{The Data}
\label{data}
\begin{figure}
\centerline{\epsfig{file=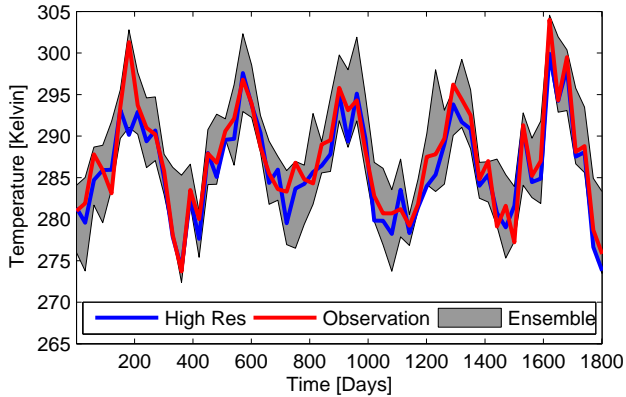,width=8.3cm}}
\caption{The Observation (red/light gray), the high resolution forecast (blue/dark gray), and the ensemble
(gray patch) over time. The entire data set comprises five years. In this
plot, all data has been down-sampled to a point every 30~days to facilitate
visualization. For the same reason, rather than showing all individual
ensemble members, the entire range of the ensembles is shown as a patch.}
\label{thedata}
\end{figure}
The forecasts and the corresponding verifications used in this study were provided by the European Center for Medium Range Weather Forecasts (ECMWF). 
The data contains the ECMWF's operational medium range deterministic forecast, run on $T_{L}799L91$ \cite{ECMWFuserguide}, to which we will refer in the following as {\sl high resolution forecast}.
Furthermore, the ECMWF's ensemble forecast $T_{L}255L40$ \cite{ECMWFuserguide} consists of 50 ensemble members, generated on $T_{L}255L40$ resolution with perturbed initial conditions, plus one member which evolves the unperturbed conditions, the {\sl control}. 
We interpret the control as an additional ensemble member; thus the total number of ensemble members is $M=51$.
Each data set consists of five subsets that correspond to four neighboring grid points on the circulation model.
The fifth data set is obtained by interpolation of the four surrounding data sets.
For all following considerations we use this interpolated data set, and we focus on the temperature forecast for London Heathrow airport, 51$^o$29'N 000$^o$27'W at noon, with a lead time of $120$h.
However, a preliminary study suggested that one could obtain qualitatively similar results for other lead times.
Fig.~\ref{thedata} shows the data set under study, including high resolution temperature forecast, ensemble forecasts and observation. 
The data covers the years from 2001 to 2005, comprising $N=1814$ data points.

In the following, $h=\{h_n\}$ denotes the time series generated by the high resolution temperature forecast issued at time $t_n=t_0 + n \Delta t$, with $n=1, \ldots, N$ and the time step $\Delta t$ being one day. 
For the data set under study the number of time instances is $N=1814$, starting on $t_0 = $1~January~2001, 12:00~UTC. 
The corresponding time series of ensemble forecasts are denoted by $x^{i}=\{x_n^{i}\}$, with $i=1,2,\ldots ,M$ referring to each ensemble member and $n$ specifying time, as above.
Analogously, the time series of verifications is denoted by $y=\{y_n\}$.
%
%
%
%%%%%%%%%%%%%%%%%%%%%%%%%%%%%%%%%%%%%%%%%%%%%%%%%%%%%%%%%%%%%%%%%
%
\section{Characterizing the Events of Interest}
\label{events}
%
%
%%%%%%%%%%%%%%%%%%%%%%%%%%%%%%%%%%%%%%%%%%%
\begin{figure}
\centerline{\epsfig{file=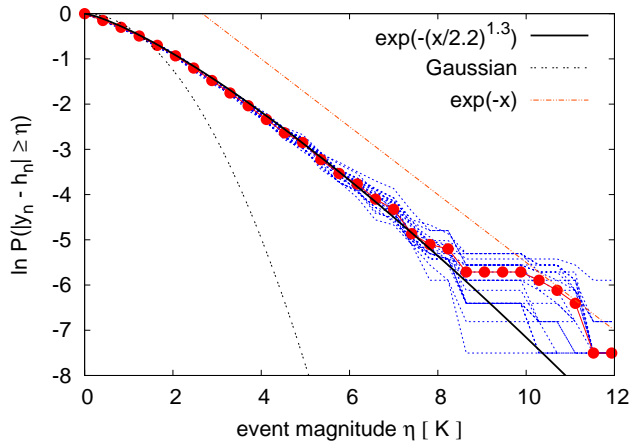,width=6cm, angle=-90}}
\caption{Probability to find events of magnitude $\eta$ in the given data set. 
The red symbols denote the probability distribution obtained from the original data set, the blue symbols indicate the probability distributions obtained from re--sampled data sets generated by drawing with repetition.
\label{ptotal}}
\end{figure}
%%%%%%%%%%%%%%%%%%%%%%%%%%%%%%%%%%%%%%%%
%
We are interested in forecasting events consisting of deviations of the high resolution forecast $h_n$ and the verification $y_n$ that exceed a given event magnitude $\eta$. 
More precisely, we define the observation variable $\chi_n(\eta)$ for the events of interest in the following way
\begin{eqnarray} 
 \chi_n(\eta) = \left\{ 
\begin{array}{c@{\quad if\quad }c} 0 & |y_n-h_n| < \eta \\
 1 & |y_n-h_n|  \geq \eta \end{array}  \right. ,  \label{weatherevent}
\end{eqnarray}
where $\chi_n(\eta) =1$ indicates an event at time step $t_n$, and $\chi_n(\eta)=0$ consequently describes the absence of an event at $t_n$.  
Hence, we obtain an additional time series $\chi(\eta)=\{\chi_n(\eta)\}$
%, with $n=1,2,\; ... \; ,N$ 
%
that keeps track of the occurrence of events.

The magnitude of an event is often measured in multiples of the standard deviation of the time series under study.
Since it is --in the context of weather forecasting-- more relevant to measure the absolute difference between a predicted temperature and the observed temperature, we prefer in this contribution to measure $\eta$ in absolute values, i.e., in Kelvin.

With respect to an analysis of extreme and rare events,
%, which are in this case large values of the time series $\{|y_n-h_n|\}$, 
it is useful to start with an overview on the range of the values and estimates
of the first moments for all relevant subsets of the multivariate time series.
\begin{center}
{\small
\begin{tabular}{|c||c|c|c|c|}
\hline 
& mean & standard & maximal & minimal \\
 &  & deviation & value & value\\ \hline
$\{y_n\}$& 287.11 K& 6.27 K &307.9K  &272.3 K \\
\hline
$\{h_n\}$& 286.57 K& 6.24 K& 305 K & 271.86 K\\
\hline
$\{x_n\}$&286.02 K &6.05 K &306.33 K &270.1 K\\ 
\hline 
$\{|y_n-h_n|\}$ & 1.98 K& 1.61 K & 12.35 K & $ 5 \times 10^{-5}$ K \\
\hline
$\{|y_n-x_n|\}$ &1.96 K & 1.71 K& 14.21 K & $3 \times 10^{-5}$ K  \\
\hline
\end{tabular}
}
\end{center}

Hence, the largest observed event $|y_n-h_n|=12.35$ K has magnitude $7.67$ times the standard deviation of the corresponding time series.
Whether it is justified to call an event {\sl extreme} if it is about $7$ times larger than the standard deviation depends on the underlying distribution. 
While a $5\sigma$ event occurs only once within a Gaussian distributed data set of $10^{7}$ i.i.d.~random numbers, one can observe $1000$ events larger then $18$ times the standard deviation in a power law distributed i.i.d.~data set of equal size \cite{Sarah3}.
The distribution of the events under study can be described reasonably well by a
 stretched exponential function as shown in Fig.~\ref{ptotal}.
In addition to the distribution estimated from the original data set, we evaluate also distributions within $20$ sample data sets of equal lengths created by drawing with repetition from the original data set.
This bootstrap method reflects the robustness of the estimated distribution towards small changes in the composition of the data set, which become especially prominent in the case of rare events. 
Extrapolating the relative frequency found in the data set under study suggests that one can expect to find about $10000$ events of size $7.7$ times $\sigma$ in a record of length $10^{7}$.
Hence the largest event we observe within the limited size of the data set is not that rare for an exponential distribution.
However, it is larger and can be expected to occur more often than the largest event one would expect if the deviations $\{|y_n-h_n|\}$ were assumed to be Gaussian distributed.

%%%%%%%%%%%%%%%%%%%%%%%%%%%%%%%%%%%%%%%%%%%%%%%%%%%%%%%%%%%%%%%%%%%%
\section{Identification of the Precursor \label{weatherprecursor}}
\label{precursor}
\begin{figure}
\centerline{\epsfig{file=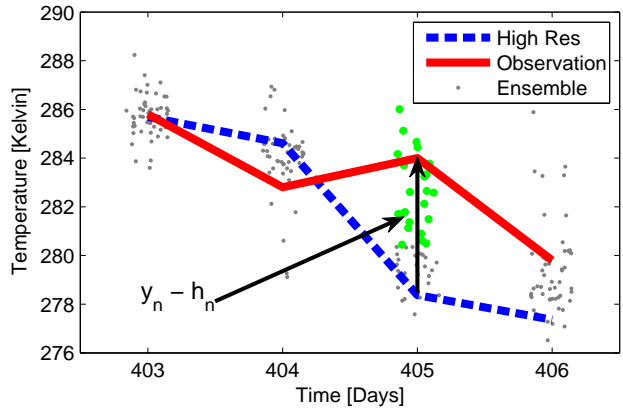,width=8.3cm}}
\caption{
\label{precvarbild}
The observation (line) and the high resolution forecast (dashed line) are shown for
four consecutive days, along with the ensemble (gray dots). In this lead
shot plot, the scattering of the ensemble members along the abscissa has no
significance and is supposed to allow for a better visualization. The
deviation $y_n - h_n$ is marked at day No.~405. Also at this day, all
ensemble members with a distance of more than 2~degrees from the High Res.\
are represented by green bold dots. There are 23~such ensemble members. Hence, the precursory variable $v_n$ has the value $v_n=23/M$.
}
\end{figure}
In this contribution we use two (of many possible) strategies to identify
precursory patterns that could announce large differences between high resolution forecast and ensemble forecast.
In both cases, we assume that useful precursory patterns can be found by investigating differences of the high resolution forecast and the corresponding ensemble forecasts.
More specifically, as precursory variable $v_n(\eta)$ we use the relative number of ensemble members that display a difference to the high resolution forecast which is larger than a specified magnitude $\beta$, i.e.,
\begin{equation}
v_n(\eta) =  \frac{\#\{i,|h_n-x_n^i| \geq \beta\}}{M} \label{wprec}.
\end{equation}
Here, $x_n^{i}$ denotes the value of the $i$-th ensemble member at time $t_n$ and $M$ is the total number of ensemble members.
Fig.~\ref{precvarbild} illustrates this definition of $v_n$.
One might consider choosing the threshold $\beta$ as a function of the event magnitude.
However, the empirical study presented in this contribution shows that even the simplest choice for $\beta(\eta)$, i.e., $\beta=\eta$ allows to make reasonable predictions.
Having defined $v_n$, we can derive a secondary variable $\hat{p}_n(v_n)$ that is then used for deciding whether to give an alarm or no alarm for an event.
We use two different strategies to define this secondary variable, namely the variable $v_n$ itself and an approach based on maximizing a conditional probability distribution function (CPDF).
Both approaches will be explained in detail in the later part of this section.
According to these strategies for decision making, 
\begin{subnumcases}{\hat{p}_n := \label{pn}} 
p(\chi_n(\eta)=1|v_n), &{CPDF method}; \label{cpdf}\\
v_n, &{counting method}. \label{counting}
\end{subnumcases}
Here, $p(\chi_n(\eta)=1|v_n)$ denotes the %conditional probability distribution 
CPDF for finding an event if a certain value of $v_n$ is observed.  
We then give an alarm for an event $\chi_n(\eta)$ to occur at a time step $t_{n}$, if the present value of $\hat{p}_n$ is larger or equal than a given threshold $\delta$.
This announcement of an alarm, based on the present value of $v_n(\eta)$ is reflected by a binary decision function
\begin{eqnarray}
A\left(\hat{p}_n, \delta\right) & = & \left\{\begin{array}{ll}
1 \quad:& \mbox{if}\; \hat{p}_n \geq \delta,\quad \mbox{with} \quad \delta \in [0,1]\\
 0 \quad:& \mbox{otherwise}.\end{array} \right. \label{decision}
\end{eqnarray}
The choice of the threshold $\delta$ reflects the tolerance towards deviation of the observable $\hat{p}_n$ from the maximum value that $\hat{p}_n$ can assume, i.e.,~unity.
As $\delta \rightarrow 1$ values of $\hat{p}_n$ that lead to an alarm are either close to the maximum of the CPDF or the maximum of the relative number of deviating ensemble members.  
Consequently small values of $\delta$, correspond to frequent alarms and lead also to a high number of false alarms.
The first method of defining $\hat{p}_n$ as introduced in Eq.~(\ref{cpdf}) is based on a maximizing $p(\chi_n(\eta) = 1|v_n)$ .
This approach corresponds to the so-called naive Bayesian classifier.
%
%%%%%%%%%%%%%%%%%%%%%%%%%%%%%%%%
\begin{figure}
\centerline{\epsfig{file=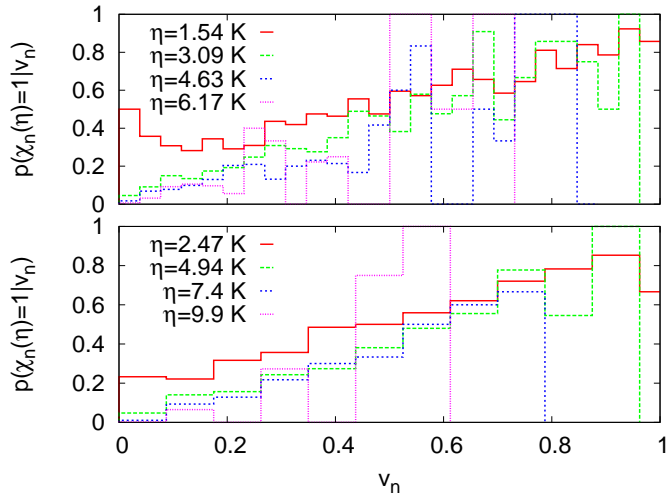,width=6.5cm, angle=-90}}
\caption{Numerical estimates of the CPDF $p(\chi_n(\eta)=1|v_n)$ as they are used for the computation of the Brier score, the ignorance and ROC-curves.
We chose to work with the number of bins $b$ that generate optimal scores, i.e., $b=12$ for the computation of the Brier score and the ignorance and $b=26$ for the computation of ROC-curves.
 \label{likelihood}}
\end{figure}
%%%%%%%%%%%%%%
%
For the numerical estimates of the CPDFs the numbers of bins are chosen with respect to the various measures for the quality of the predictions that will be introduced in the following section.
In more detail, the number of bins $b$ is chosen such that we observe the respective score to be optimal in the regime where sufficiently many events are available, i.e., for small and intermediate values of $\eta$. 
Fig.~\ref{likelihood} shows examples for the estimates of the CPDF used for the generation of Brier scores and ignorance scores ($b=12$ in both cases) and for the generation of ROC-curves (b=26).
Using more bins increases the specificity and hence improves the resulting ROC~curves. 
On the other hand, increasing the number of bins leads to an increased variance of the estimated CPDFs, especially in the limit of very few events, i.e., large values of $\eta$.
We test for this increase in variance through cross-validation.

The second method of identifying suitable precursory structures is based on the assumption that the ensemble members represent equally likely scenarios of the future evolution of the atmosphere.
Hence a large ensemble spread, reflects a state of the atmosphere in which forecasts are difficult to make.
In other words, if the high resolution forecast differs significantly from the verification, we assume that this failure of the forecast is due to an increased sensitivity on small perturbations in the initial conditions. 
We can then hope that this sensitivity is not only present in the atmosphere, but also reflected by the ECMWF's circulation model.
Consequently, we would expect a large number of ensemble members to deviate from the high resolution forecast.
% that is then turned into an alarm for a deviation.
%

The estimates of the CPDFs, as shown in Fig.~\ref{likelihood} indicate that we can assume $p(\chi_n(\eta)=1|v_n)$ to be a monotonous function (if we attribute the fluctuations to finite sample effects).
Consequently, the values of $v_n$ that lead to an alarm for an extreme event in terms of the counting approach will in many cases also generate an alarm in the CPDF-approach.
Remembering that $p(\chi_n(\eta)=1|v_n)$ are the observed frequencies of events, given $v_n$ and interpreting $v_n$ as the forecast probabilities for an event, we can even think of the plots in Fig.~\ref{likelihood} in terms of a reliability diagram \cite{MurphyWinkler1977}. 
If the original ensemble $x_n^i$ and the verification $y_n$ were independent draws from the same distribution, the curves should coincide with the diagonal.
For very large and very small events, we observe deviations from the diagonal, that can be either attributed to the limited amount of available data in these regimes or to systematic deviations. 
%

%
%%%%%%%%%%%%%%%%%%%%%%%%%%%%%%%%%%%%%%%%%%%%%%%%%%%%%%%%%%%%%%%5
\section{Evaluating the Quality of the Predictions}
%
%%%%%%%%%%%%%%%%%%%%%%%%%%%%
\begin{figure}
\centerline{\epsfig{file=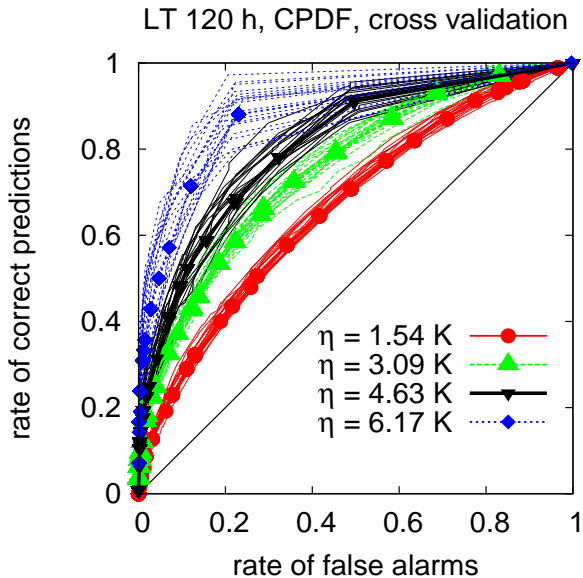,width=8cm, angle=-90}}
\caption{ROC-curves generated through estimates of conditional probability distributions, i.e $\hat{p}_n=p(\chi_n(\eta)=1|v_n)$., and leave-one-out cross-validation.
As in Fig.~\ref{roc1}, the lines without symbols represent ROC-curves computed within re-sampled data sets.
\label{roc2}
}
\end{figure}
%%%%%%%%%%%%%%%%%
%
\label{quality}
In the following sections we present the results for the predictions made according to the CPDF method and the counting method as specified in Sec.~\ref{weatherprecursor}.
In order to evaluate the quality of the predictions we use different measures, namely the ROC-curve \cite{Egan} and the Brier Score \cite{Brier} and the Ignorance \cite{roulston}. 
In-sample predictions were made for the lead times 24h, 48h, 96h, 144h, 168h, 192h, 216h, 240h.
Since we do not find qualitatively different results for different lead times (with the exception of 24h) we restrict ourselves to the discussion and further investigation of the forecasts with a lead time of 120h. 
To support the validity of the results obtained from our relatively small sample (1814 data points) we have to consider different sources of uncertainty:
the robustness of our results towards small changes in the specific composition of the data set and the influence of over-fitting due to in-sample prediction.
The later effect is only an issue for the CPDF method, since no training is needed to identify the precursor for the counting method.
 We test for the robustness towards small changes in the composition of the sample by re-sampling the data set (Bootstrap method). 
The re-sampling is done by creating 20 test data sets by drawing with repetition from the original data set and applying the same training and prediction algorithm.
Using the re-sampled data sets leads to a slightly different estimate of the probability distribution.
Additionally to the results obtained from the original data set, we hence obtain a distribution of results based on the re-sampled data sets, which can serve as an estimate of the variance of the original results.
The random generator used for the re-sampling is the {\sl Mersenne twister} \cite{Mersennetwister}, as implemented in the {\sl Gnu Scientific Library} \cite{gsl}.
In order to test for the effects of in-sample prediction, we repeat the predictions doing leave-one-out cross validation (also called {\sl total cross validation}). 
We therefore train on all but one data points and predict the occurrence/non-occurrence of an event in the left out time step.
In the context of maximum CPDF estimation, ``training'' refers to determining the CPDF.
This procedure (training and prediction) is repeated $N$ times, with $N$ being the number of data and a different data point left out for each repetition.
Since the counting method needs no training, we apply the leave-one out cross validation only for the CPDF method.
%
%%%%%%%%%%%%%%%%%%%%%%%%
\subsection{ROC and AUC} \label{rocsection}
%%%%%%%%%%%%%%%%%%%%%
\begin{figure}
\centerline{\epsfig{file=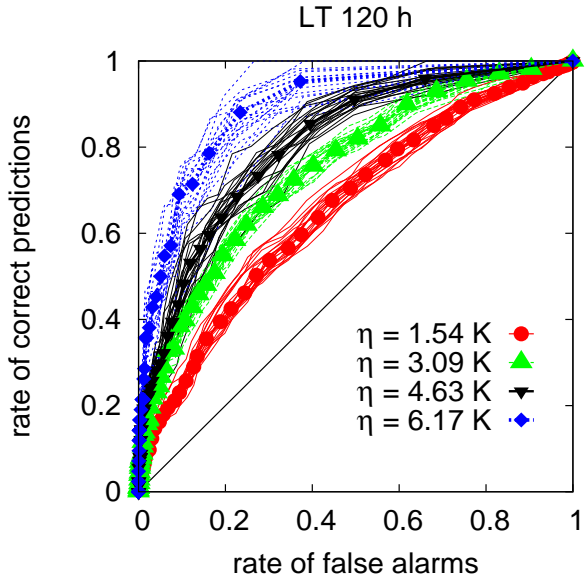,width=8cm, angle=-90}}
\caption{ROC-curves for the prediction of differences between a high resolution forecast and its verification. 
The ROC-curves were created using the number of deviating ensemble members as a predictor, i.e., $\hat{p}_n=v_n$.
This method is introduced as counting method in Sec.~\ref{weatherprecursor}.
The points represent the original data set, the lines represent 20 bootstrap-samples. \label{roc1}}
\end{figure}
A common method to evaluate the success of a classification task is the receiver operating characteristic curve (ROC-curve) \cite{Swets1, Egan}.
We first compute the rate $r_{c}$ of correctly
predicted events {\sl (hit rate, rate of true positives, sensitivity)} to the rate $r_{f}$ of false alarms {\sl(rate of false positives, 1$-$specificity)}.
A ROC-curve comprises a plot of $r_c$ against $r_f$ as e.g.~in Figs.~\ref{roc1} and \ref{roc2}. 
Numerically, these rates can be computed from the time series of the precursory variable $\{v_n\}$ and the time series of the events $\{\chi_n(\eta) \}$ by simple counting.
For each value of the threshold $\delta$ one obtains a point in the $r_c$--$r_f$ plane.
 If $\delta$ is assumed to be a continuous variable one arrives at a curve parametrized by $\delta$.
The resulting curve in the unit-square of the $r_f$-$r_c$
plane  approaches the origin for $\delta \rightarrow 1$ and the point $(1,1)$
in the limit $\delta \rightarrow 0$. 
A curve above the diagonal reveals that the corresponding strategy of prediction is
better than a purely random prediction, which is characterized by a ROC-curve along the
diagonal.
%
%
%%%%%%%%%%%%%%%
\begin{figure}
\epsfig{file=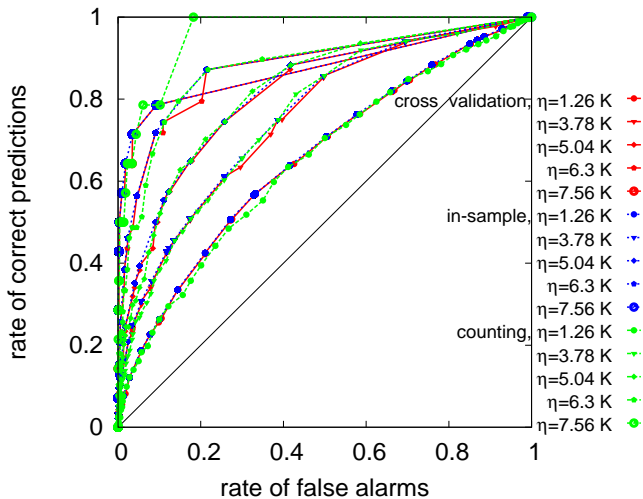,width=6.8cm, angle=-90}
\caption{Comparison of counting and CPDF method.
\label{roc3}
}
\end{figure}
%%%%%%%%%%%%%%%%%
% 

\begin{figure}
\centerline{\epsfig{file=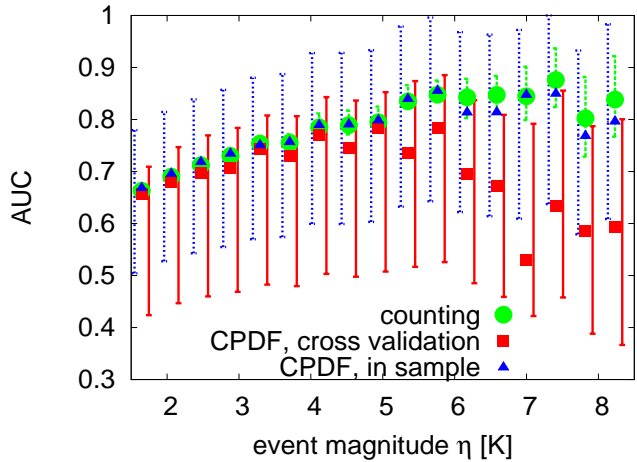,width=6.3cm, angle=-90}}
\caption{The area under the ROC-curves for the predictions according to maximum  CPDF and counting method. 
The two ROC-curves for the CPDF method correspond to in-sample predictions and leave-one-out cross validation.
Confidence-bars were evaluated using re-sampled data sets, as it is described in more detail for the generation of confidence-bars for scores in Sec.~\ref{bandi}.
 \label{auc}}
\end{figure}

ROC-curves that characterize the success of counting method and CPDF method are presented in Figs.~\ref{roc1} and \ref{roc2}.
The lines with symbols represent ROC-curves generated from the original data set. 
The lines without symbols represents ROC-curves computed from 20 re-sampled data sets.
As mentioned in the previous sections the number of bins used to estimated the CPDF was chosen with respect to the methods used to quantify the success of the predictions.
Concerning ROC-curves, we found that a finer binning lead to optimal results, i.e, the number of bins used to estimate CPDFs is $26$.
Although the corresponding estimates of conditional probability distributions are not smooth functions of $v_n$, they produce reasonable good ROC-curves, since the fine binning results in a high specificity of the observed precursor.
One can see that the quality of the forecasts (as far as it is quantified by the ROC curve) increases with the magnitudes of the events under study. 
This result is consistent with the findings previously observed for precursor based predictions in time series \cite{Sarah4} and to the prediction of precipitation \cite{Goeber}. 
If we focus on larger events, then also the sensitivity to small variations in the data set under study increases, as is shown by the spread of the ROC-curves obtained from the re-sampled data sets. 
This is not surprising, since larger events occur less often and hence small deviations in their frequency of occurrence become more prominent. 
Figs.~\ref{roc3} and \ref{auc} compare CPDF and counting method, as well as the influence of the cross validation.
The area under the ROC-curve (AUC) is a well established summary index for ROC-curves, see e.g., \cite{Pepe} for other summary indices.
An optimal prediction is characterized by an AUC of unity, random predictions are reflected by a diagonal in the ROC-plane and hence an AUC of $0.5$.

Both the ROC-curves in Fig.~\ref{roc3} and the corresponding AUCs in Fig.~\ref{auc} show that the success of counting method and CPDF method does not differ significantly in the regime of $\eta \leq 5$.
%
%%%%%%%%%%%%%%%%%%%%%%%%%%%%
%%%%%%%%%%%%%%%%%
%
For larger event magnitudes, i.e., in the regime of very few events, the counting method performs better than the CPDF method. 
In this regime, the difference between in-sample prediction and leave-one-out cross validation becomes more prominent as well.
Comparing the AUCs in Fig.~\ref{auc} with the scores in Figs.~\ref{briercv} and \ref{ignocv} in the following chapter, the regime where one can observe a clear difference between counting method, in-sample prediction, and total cross-validation starts earlier, that is for smaller values of $\eta$.
This observation can probably be attributed to a higher variance of the estimates of the conditional probability distribution used to generate AUCs and ROC-curves, provoked by an increased number of bins. 
In theory, choosing $\hat{p}_n=p(\chi_n(\eta)=1|v_n)$ should be the optimal strategy for a prediction, if it could be estimated with arbitrary accuracy. 
However, as we saw in Figs.~\ref{roc1}-\ref{auc}, given a finite data set, the counting method leads to very similar or even better results. 
The success of the counting method can be understood by taking into account that the CPDF (as indicated by Fig.~\ref{likelihood}) appears to be monotonously increasing with $v_n$.
Hence, 
the maximum of the CPDF can be expected to be close to large values of $v_n$  that are also considered to be good precursors in terms of the counting approach.
Or vice versa, a value of $v_n$ that leads to an alarm according to the counting method, does also lead to an alarm in terms the maximum CPDF approach.
However, since the counting method is not based on any training procedure or method of estimation it is independent on the number of available events.
Consequently, the counting methods performs very similar to the CPDF approach, if the CPDF can be estimated well from the available number of events,
 and it leads to better predictions, if the estimates of the CPDF are poor, as e.g., in the case of very large events.
%
%
%%%%%%%%%%%%%%%%%%%%%%%%%%%%%%%%%%%%%%%%%%%%%%%%%%%%%%%%%%%%%%%%%%%%%%%%%%%
%
%%%%%%%%%%%%%%%%%%%%%%%%%%%%%%%%%%%%%%%%%%%%%%%%%%%%%%%%%%%%%%%%%%%%%%%%%%%
\subsection{Brier Score and Ignorance}
\label{bandi}
To test whether the results presented in the previous section depend on a measure for the quality of a prediction, we also compute Brier scores \cite{Brier} and ignorance scores \cite{roulston}. 
Both scores are common methods to test for qualities of predictions in the framework of weather forecasting.
The Brier score \cite{Brier} is defined as 
\begin{eqnarray}
B\left(\chi(\eta),\hat{p}\right) & = & \frac{1}{N} \sum_{n = 1}^N (\chi_n(\eta) - \hat{p}_n)^2,
\label{brier}
\end{eqnarray}
where $\hat{p}=\{\hat{p}_n\}$ denotes the time series of successive $\hat{p}_n$.
As defined in Eq.~(\ref{pn}) $\hat{p}$ is either chosen to be the CPDF $p\left(\chi_n(\eta)=1| v_n\right)$ or the relative number of ensemble members $v_n(\eta)$.
A suitable choice of $\hat{p}_n$, produces a small value of $B\left(\chi(\eta),\hat{p}\right)$, since we expect the value of $\hat{p}_n$ to be approximately unity if an event is observed ($\chi_n(\eta)=1$) and to be close to zero otherwise.
%
%%%%%%%%%%%%%%%%%%%%%%%%%%
\begin{figure}
\centerline{\epsfig{file=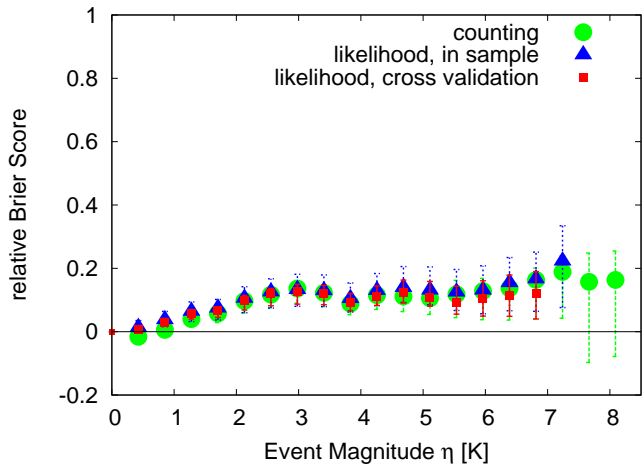,angle=-90, width=9cm}}
\caption{Relative Brier scores for the counting method and the maximum CPDF approach.
In order to test for over-fitting, the CPDFs were additionally estimated by
 using leave-one-out-cross-validation.
The generation of the confidence-bars is explained in more detail in Sec.~\ref{bandi}.
\label{briercv}}
\end{figure}
%%%%%%%%%%%%%%%%%%%%%%%
%
Relative scores (Skill scores) measure whether a forecast method is better than the forecast given by the relative frequency $f=p\left(\chi_n(\eta)=1\right)$ of events.
Consequently, the relative Brier score (Brier skill score) is defined as
\begin{equation}
B_{rel}\left(\chi(\eta), \hat{p}, f\right) = \frac{B\left(\chi(\eta), f\right) - B\left(\chi(\eta), \hat{p}\right)}{B\left(\chi(\eta), f\right)},
\end{equation}
where $B\left(\chi(\eta), f\right)$ denotes the Brier score obtained for the relative frequency of observed events $f$.

The ignorance score \cite{roulston} for a binary event is given by
\begin{eqnarray}
I\left(\chi(\eta),\hat{p}\right) & = & \frac{1}{N} \sum_{n = 1}^N -\log(\hat{p}_n) \cdot \chi_n(\eta) - \nonumber\\
&{\it }&\quad \quad \quad \log(1 - \hat{p}_n) \cdot (1 - \chi_n(\eta)).
\label{igno}
\end{eqnarray}
Similar to the Brier score, a good prediction is characterized by a small value of the ignorance score. In other words, one aims at minimizing the ignorance.

To make sure that numerical estimates of $\hat{p}_n$ do not cause a divergent ignorance, we add two imaginary ensemble members.
Of these two imaginary members, one is always counted as a deviation from the high resolution, while the second one is assumed to never deviate from the high resolution forecast.   
Thus, we always ensure that $\frac{1}{M+2} \leq \hat{p}_n \leq \frac{M+1}{M+2}$.
This method of regularization introduces some bias towards the estimation of $\hat{p}_n$, but it is necessary in order to use the ignorance for the evaluation of yes/no-forecasts (binary forecasts).
Analogously to the relative Brier score, the {\sl relative ignorance} 
\begin{equation}
I_{rel}\left(\chi(\eta), \hat{p}, f\right) = \frac{I\left(\chi(\eta), f\right) - I\left(\chi(\eta), \hat{p}\right)}{I\left(\chi(\eta), f\right)}
\end{equation}
compares the ignorance obtained from the predictive distribution $\hat{p}$ with the ignorance of the climatology $f$.
Both, Brier score and ignorance indicate a good quality of the predictions when their values are small.
However, this implies that the respective relative scores are expected to approach unity, for an ideal forecast and zero for a forecast, that is not better than the climatology. 
In other words, when comparing relative scores, a larger value of a relative score indicates a better forecast.
%
%%%%%%%%%%%%%%%%
\begin{figure}
\centerline{\epsfig{file=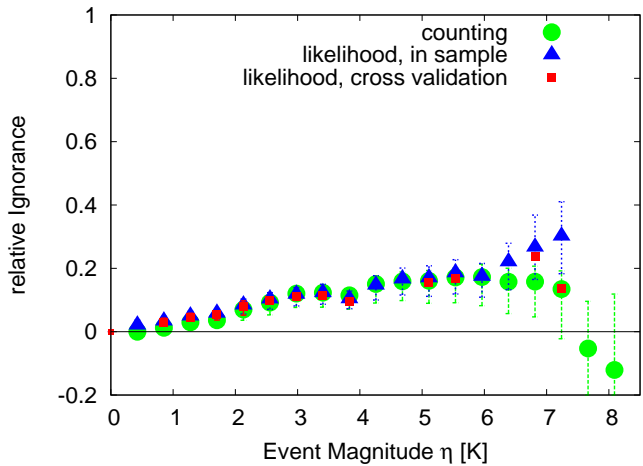, angle=-90, width=9cm}}
\caption{The relative ignorance scores of the decision making strategies defined in Eq.~(\ref{pn}). 
In order to test for over-fitting, the estimation of the conditional probability density was repeated with leave-one-out-cross validation.
The generation of confidence-bars is explained in more detail in Sec.~\ref{bandi}.
\label{ignocv}}
\end{figure}
%%%%%%%%%%%%%%%%%%

The relative scores evaluated for counting and CPDF method are shown in Figs.~\ref{briercv} and \ref{ignocv}. 
The confidence bars in Figs.~\ref{briercv} and \ref{ignocv} reflect the variance of the ensemble of re-sampled data sets.
Especially for large event magnitudes, not all re-sampled data sets provide a sufficient number of events for the evaluation of scores.
Hence, if all re-sampled data sets allowed the computation of scores, the confidence bars were estimated from $20$ re-sampled data sets. 
However, if the re-sampled data sets did not provide estimates of the scores, for given values of $\eta$ and $v_n$, we estimated the confidence bars from a smaller number of re-sampled data sets.
We decided to plot an estimated confidence bar for any given values of $\eta$ and $v_n$ whenever at least 5 re-sampled data sets provided scores.
Both, Brier score and the ignorance increase with increasing event magnitude in the regime $0\leq \eta \leq 3$. 
For $\eta \geq 3$ the Brier Score is approximately constant, while the ignorance continues to increase until $\eta \approx 6$.
If $\eta \geq 6$, the confidence bars of both scores increase significantly, and for the ignorance we observe in addition a qualitatively different dependence on $\eta$ for counting and CPDF method. 
Both effects can probably be attributed to finite sample effects that become more prominent for larger and thus rarer events. 

With respect to the comparison of the two different strategies of defining $\hat{p}_n$, the scores show results that are consistent with the ROC and AUC~curves discussed in the previous section.
For small values of $\eta$, i.e., in the regime where many events are available and the uncertainty in the estimation of the scores is small, the results for the counting method and the CPDF method with and without cross validation coincide.
The increase of the difference between in-sample CPDF prediction and cross-validated CPDF prediction can be understood by considering the effect of leave-one-out-cross-validation on scores.
By reordering the temporal indices in the summation in Eq.~(\ref{brier}) and ~(\ref{igno}) one arrives at the following expressions for scores $B_{CV}$ and $I_{CV}$ estimated with leave-one-out-cross-validation (CV)
\begin{eqnarray}
B_{CV} &=& \frac{1}{N}\sum_{k}  N_k \frac{N_k^2}{(N_k-1)^2} \; \hat{p}_k \left(1-\hat{p}_k \right), \label{Bloo}\\
I_{CV} &=& \frac{-1}{N} \sum_{k} \left[ \hat{p}_k \ln \left(\frac{M_{k}-1}{N_{k}-1}\right)  \right. \nonumber \\
&{\it}& \quad \quad \quad \quad \left. +\left(1 - \hat{p}_k \right)\ln \left( 1 - \frac{M_k}{N_{k}-1}\right)\right]
\end{eqnarray}
where $N_k$ denotes the number of entries in the bin associated with the value $v_k$ and $M_k$ the number of entries in bin $k$ that coincide with an event. 
According to Eq.~(\ref{Bloo}), the terms that contribute to Brier scores estimated with leave-one-out-cross-validation are by a factor of $N_k^2/(N_k -1)^2$ larger than the corresponding terms evaluated on the full data set.
Since $M_k \leq N_k$, it is easy to see that the ignorance evaluated by leave-one-out-cross-validation is also larger than the ignorance computed on the whole data set. 
Consequently one can expect the corresponding relative scores to be smaller.
For both scores the effect of the cross-validation becomes more prominent in the limit of very few entries in the histogram of the CPDF,  i.e., small values of $N_k$ as we observe for very large values of $\eta$.

%%%%%%%%%%%%%%%%%%%%%%%%%%%%%%%%%%%%%%%%%%%%%%%%%%%%%%%%%%%%%%%%%%%%%%%%%
\section{Understanding the Dependence on the Event Magnitude \label{condition}}
%%%%%%%%%%%%%%%%%%%%%%%%
\begin{figure}
\centerline{
\epsfig{file=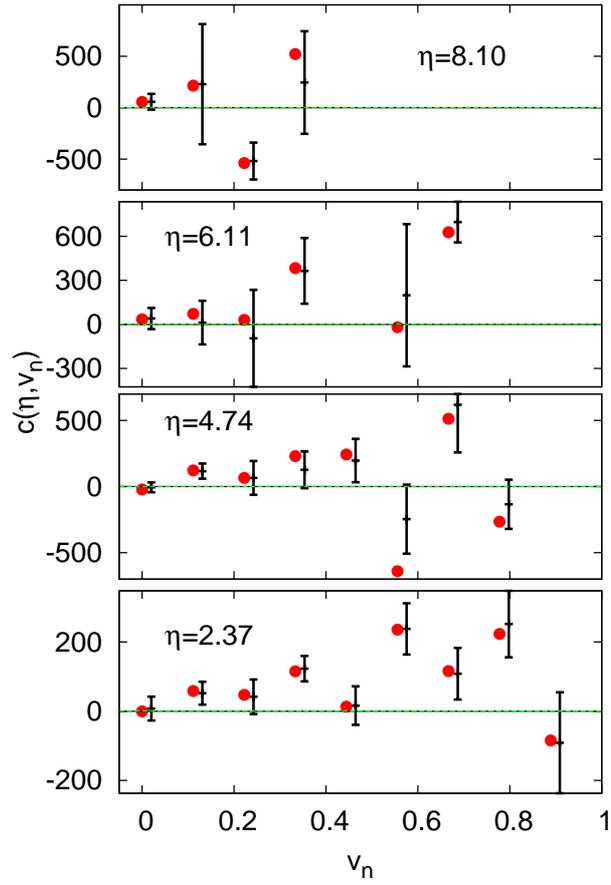, width=8.5cm}
}
\caption{
The numerical estimates of $c(\eta,v_n)$ according to Eq.~\ref{c2b}, as a function of $\rho$ for selected fixed values of $\eta$; confidence bars were again created using $20$ re-sampled data sets and drawn if at least $5$ of the $20$ data sets provided estimates of $c(\eta,v_n)$.  \label{weathercondi2}}
\end{figure}
\label{magnitude}
The ROC-curves and scores presented in the previous section indicated a better predictability of larger events. 
Intuitively one could expect these larger events to be harder to predict, since they are rare.
On the other hand, they are clearly distinguishable from the average event, and one can thus expect that they are also preceded by a more distinguished precursory signal.

\begin{figure}
\centerline{\epsfig{file=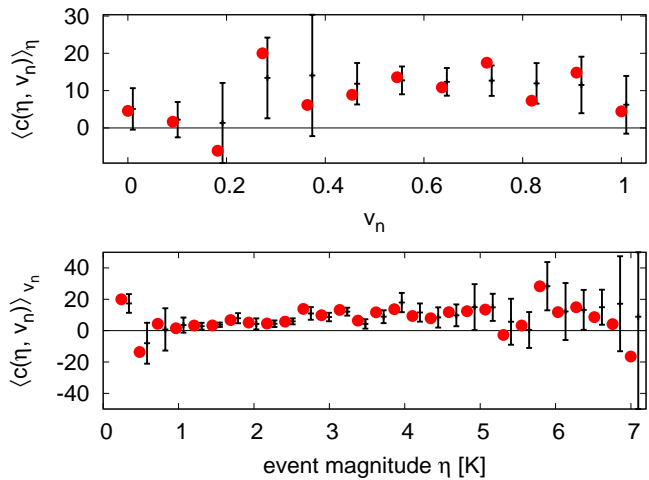, angle=-90, width=9cm}}
\caption{The lower figure shows the condition $c(\eta, v_n)$, averaged over all investigated values of $v_n$. 
The upper graph displays the condition $c(\eta, v_n)$, averaged over all investigated values of $\eta$. 
\label{weathercondiav}}
\end{figure}

In order to test for the magnitude dependence, we consider again ROC-curves as a measure for the quality of a prediction. 
The slope of the ROC-curve can be identified to be the likelihood ratio (\cite{Egan})
\begin{equation}
\Lambda\left(v_n, \eta\right) = \frac{p\left(v_n|\chi_n(\eta)=1\right)}{p\left(v_n|\chi_n(\eta)=0\right)}. \label{lr}
\end{equation}
If we understand the likelihood ratio as a function of $v_n$, Eq.~(\ref{lr}) describes a family of functions parametrized by $\eta$. 
We can then simply investigate the dependence of the likelihood ratio on the event magnitude
by computing the derivative of $\Lambda\left(v_n,\eta\right)$ with respect to the event magnitude. 
Rearranging the equation yields the following condition \cite{Sarah2}, 
\begin{eqnarray}
c(\eta, v_n) &=& \frac{\partial}{\partial \eta}\ln p\left(\chi_n(\eta)=1|v_{n}\right)- \nonumber \\
&{\it}&
\frac{\bigl(1-p\left(\chi_n(\eta)=1|v_{n}\right))\bigr)}{\bigl(1-p\left(\chi_n(\eta)=1\right)\bigr)}\;\frac{\partial}{\partial
\eta}\ln p\left(\chi_n(\eta)=1\right) \label{c2b}.
\end{eqnarray}
This expression has the same sign as the derivative of the likelihood ratio, i.e.,
\begin{equation}
\mbox{sign} \left(
\frac{\partial}{\partial \eta} \Lambda(v_n,\eta)\right) = \mbox{sign} \;c(\eta,v_n). 
\end{equation}
Consequently, if  $c(\eta,v_{n})> 0$, the corresponding families of likelihood ratios and ROC-curves reveal that the quality of the predictions increases, if one focuses on larger events.
If $c(\eta,v_{n})< 0$, the corresponding families of likelihood ratios and ROC-curves show a negative magnitude dependence. In this case larger events are harder to predict, and if $c(\eta,v_{n})=0$, likelihood ratio and ROC-curves are not dependent on the event magnitude.
The results of the numerical estimation of $c(\eta,v_n)$ are presented in Fig.~\ref{weathercondi2}.
%
%%%%%%%%%%%%%%%%%%%%%%%%%%%%%%%%%%%%%%%%%%%%%%%%%%%%%%%%%%%%%%%
\begin{figure}
\centerline{
\epsfig{file=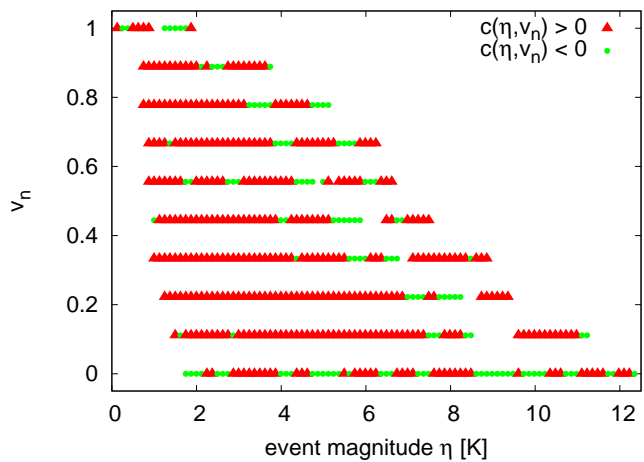, angle=-90, width=9cm}
}
\caption{This plot indicates whether $c(\eta,v_n)$ is positive or negative for a given pair of coordinates in the $v_n,\eta$ plane. 
\label{weathercondi1}}
\end{figure}
%%%%%%%%%%%%%%%%%%%%%%%
%%
%
Note that $c(\eta,v_n)$ is not dependent on the choice of the precursor and on the methods of decision making (e.g., maximum CPDF approach or counting), but simply a function of $c(\eta,v_n)$.
Nevertheless, the computation of $c(\eta,v_n)$ requires numerical estimates of the CPDF and the climatology, as well as their derivatives.
The derivatives of the estimates of both distributions are obtained using Savitzky-Golay-filtering \cite{Savgol}.
As in the previous sections, confidence bars represent the double standard deviations of results obtained from $20$ re-sampled data sets. 
However, especially large events cannot be expected to be well represented in some of the sample data sets and consequently it was for certain values of $\eta$ and $v_n$ not possible to evaluate the condition for all $20$ re-sampled data set. 
In order to use as much of the available information as possible, we decided to plot confidence bars, if at least $5$ members of the re-sampled data sets could produce estimates of $c(\eta,v_n)$.
Since the evaluation of $c(\eta,v_n)$ is based on the numerical estimation of derivatives of PDFs, finite sample effects become even more prominent, compared to the computation of the scores.
Consequently, due to the relatively small size of the data set, $c(\eta,v_n)$ is very noisy, (see Fig.\ref{weathercondi2}) compared to previous results for other prediction tasks \cite{Sarah2} that were generated using larger data sets.
Nevertheless, we find that the numerical estimates of $c(\eta,v_n)$ have positive values for the majority of coordinates in the $\eta$-$v_n$-plane, as it is shown in Fig.\ref{weathercondi1}.
% has through a wide range positive values.
%
Strictly speaking, the dependence on the event magnitude can vary for every point in the $\eta,v_n$-plane, since $c(\eta, v_n)$ is a function of both variables.
However, for practical considerations, one might be interested in the overall behavior, independently of the specific choice of the precursor or the event threshold.
That is why we also studied whether averages of $c(\eta,v_n)$, i.e, $\langle c(\eta,v_n)\rangle_{\eta}$  and  $\langle c(\eta,v_n)\rangle_{v_n}$ can also correctly characterize the overall dependence on the event magnitude, without referring to a specific value of the precursory variable or the event size.
Fig.~\ref{weathercondiav} shows that $\langle c(\eta,v_n)\rangle_{\eta}$ is positive for most values of the precursory variable $v_n$. 
In particular it is positive for larger values of $v_n$, which are relevant as precursors for large events. 
%%%%%%%%%%%%%%%%%%%%%%%
%
The averages over the precursory variable $v_n$, displayed in Fig.~\ref{weathercondiav} are also positive for almost all values of $\eta$. 
The fact that the exceptions occur within the regimes of very small and very large values of $\eta$ support the assumption, that we can attribute them to finite sample effects, since in both regimes the corresponding events are rare.
In total the evaluation of the condition $c(\eta, v_n)$ and its averages over $\eta$ or $v_n$ suggest a positive magnitude dependence.
%
%
%%%%%%%%%%%%%%%%%%%%%%%%%%%%%%%%%%%%%%%%%%%%%%%%%%%%%%%%%%%%%%%%%%%%%%%%%%%%%%%
%
%%%%%%%%%%%%%%%%%%%%%%%%%%%%%%%%%%%%%%%%%%%%%%%%%%%%%%%%%%%%%%%%%%%%%%%%%%%%%%%%
%
%%%%%%%%%%%%%%%%%%%%%%%%%%%%%%%%%%%%%%%%%%%%%%%%%%%%%%%%%%%%%%%%%%%%%%%%%%%%%%%
\section{Conclusions}
\label{conclusions}
We use information obtained from the ensemble forecast and the high resolution forecast to successfully predict errors of the high resolution forecast.
The output of the ensemble forecast system is post-processed in terms of a decision making approach.
More precisely, 
the number of ensemble members that display a large difference to the corresponding high resolution forecast can serve to predict the events we are interested in.
The quality of the predictions was evaluated using the Brier score, the ignorance score, and ROC-curves.
Comparing two different strategies of decision making, we find no significant
difference in the success of the counting method and the numerically more expensive, but theoretically better justified maximum CPDF approach.
This is surprising, since the counting method consists simply in imposing a threshold to the number of ensemble members that show large deviations from the high resolution forecast.
However, the similarity of the results can be understood if we consider the fact that the CPDF is a monotonously increasing function and consequently, the counting approach mimics the choices of suitable values of the precursory variable that would have been chosen according to the maximum CPDF approach.
Additionally, all different methods to evaluate the quality of the predictions (ROC-curves, the relative Brier score and the relative ignorance score) display an increase in the quality of the predictions with increasing event magnitude.
This increase is particularly apparent in regimes of the event magnitude that are well supported by the amount of observed events.
This positive magnitude dependence of the ROC-curves could be reproduced as well through a test condition for the magnitude dependence of ROC-curve and likelihood ratio.
%
%\begin{acknowledgements}
\section{Acknowledgements}
The authors are grateful to Renate Hagedorn and the European Centre for
Medium Range Weather Forecasting for kindly providing ensemble forecast as
well as station data for London Heathrow.
Sarah Hallerberg is grateful to the German Academic Exchange Service (DAAD) for the financial support of her stay at the Centre for the Analysis of Time Series at the London School of Economics.
%\end{acknowledgements}
%
%%%%%%%%%%%%%%%%%%%%%%%%%%%%%%
% 
\bibliographystyle{plain}
\bibliography{/home/shallerberg/artikel/complete}
\end{document}